\documentclass[twocolumn,abs,prb,showpacs]{revtex4-1}

\usepackage{graphicx}
\usepackage{dcolumn}
\usepackage{float}
\usepackage{amsmath}
\usepackage{bm}
\usepackage[mathscr]{euscript}

\newcommand{\comment}[1]{}

\begin{document}


\title{Optical Conductivity of Weyl Semimetals and Signatures of the Gapped Semimetal Phase Transition}

\author{C. J. Tabert$^{1,2}$}
\author{J. P. Carbotte$^{3,4}$}
\affiliation{$^1$Department of Physics, University of Guelph,
Guelph, Ontario, Canada N1G 2W1} 
\affiliation{$^2$Guelph-Waterloo Physics Institute, University of Guelph, Guelph, Ontario, Canada N1G 2W1}
\affiliation{$^3$Department of Physics, McMaster University,
Hamilton, Ontario, Canada L8S 4M1} 
\affiliation{$^4$Canadian Institute for Advanced Research, Toronto, Ontario, Canada M5G 1Z8}
\date{\today}

\begin{abstract}
{The interband optical response of a three-dimensional Dirac cone is linear in photon energy ($\Omega$).  Here, we study the evolution of the interband response within a model Hamiltonian which contains Dirac, Weyl and gapped semimetal phases.  In the pure Dirac case, a single linear dependence is observed, while in the Weyl phase, we find two quasilinear regions with different slopes.  These regions are also distinct from the large-$\Omega$ dependence. As the boundary between the Weyl (WSM) and gapped phases is approached, the slope of the low-$\Omega$ response increases, while the photon-energy range over which it applies decreases.  At the phase boundary, a square root behaviour is obtained which is followed by a gapped response in the gapped semimetal phase. The density of states parallels these behaviours with the linear law replaced by quadratic behaviour in the WSM phase and the square root dependence at the phase boundary changed to $|\omega|^{3/2}$.  The optical spectral weight under the intraband (Drude) response at low temperature ($T$) and/or small chemical potential ($\mu$) is found to change from $T^2$ ($\mu^2$) in the WSM phase to $T^{3/2}$ ($|\mu|^{3/2}$) at the phase boundary.
}
\end{abstract}

\pacs{78.20.-e
} 

\maketitle

\section{Introduction}

Dirac and Weyl semimetals are a class of three-dimensional (3D) systems which posses nontrivial topology\cite{Wang:2012a,Wang:2013a,Weng:2014,Weng:2015}. Recently, such materials have been studied by angular-resolved photoemission spectroscopy and scanning tunnelling microscopy\cite{Neupane:2014,Borisenko:2014,Jeon:2014,Liu:2014a,Liu:2014,Xu:2013,Lv:2015,Xu:2015a,Lv:2015b,Xu:2015d,Yang:2015a, Xu:2015b,Borisenko:2015}.   By breaking time-reversal symmetry or spatial-inversion symmetry, the double degeneracy of a Dirac node is broken and two Weyl nodes emerge.  These are separated in either momentum or energy space\cite{Vazifeh:2013}.  The simplest Hamiltonian which captures this physics describes a semimetal with a 3D electronic band structure where the energy is directly proportional to the crystal momentum via an isotropic Fermi velocity $v_F$.  More specifically, $\hat{H}_0=\hbar v_F\bm{\sigma}\cdot\bm{k}$, with $\bm{\sigma}=(\hat{\sigma}_x,\hat{\sigma}_y,\hat{\sigma}_z)$, the usual vector of Pauli-spin matrices.  

Such a Hamiltonian leads to an ac optical interband response which is linear in photon energy\cite{Timusk:2013,Ashby:2014} ($\Omega$) and passes through the origin in the limit $\Omega\rightarrow 0$.  This unusual linear behaviour is widely considered to be an important optical signature of 3D Dirac materials\cite{Xu:2015} and represents the generalization of the well-known constant interband background of graphene and similar two-dimensional (2D) materials\cite{Gusynin:2006a,Gusynin:2007,Gusynin:2009,Carbotte:2010,Li:2008,Stille:2012}.  In 2D, the background is universal, independent of any material parameters while in 3D, the linear background is inversely proportional to the Fermi velocity. An early observation of the linear optical background was that of Chen \emph{et al.}\cite{Chen:2015} in ZrTe$_5$ where the linear regime was found to extend over $\sim 150$ meV.  Furthermore, the pyrochlore Eu$_2$Ir$_2$O$_7$ displays a linear optical conductivity which extrapolates to the origin\cite{Sushkov:2015}; however, it exists over a much smaller range of frequency ($\sim 12$ meV).  The recent work of Xu \emph{et al.}\cite{Xu:2015} on the Weyl semimetal (WSM) TaAs shows two linear regions: one below $\sim 30$ meV which extrapolates to the origin, and a second which extends over $\sim 125$ meV with a much reduced slope.  Before these recent studies, Timusk \emph{et al.}\cite{Timusk:2013} considered older optical data in some known quasicrystals and provided a possible interpretation in terms of Dirac or Weyl points.  All the materials they considered had a linear conductivity with some extrapolating to a positive intercept on the conductivity axis while others cut the photon axis at $\Omega>0$.  Some showed two linear regions.  It is clear from such data that the optical properties of 3D Dirac materials display a rich pattern of behaviour and such studies are expected to provide important information about the electronic properties of these semimetals.

While TaAs and NbAs are WSMs with broken spacial-inversion symmetry\cite{Weng:2015}, YbMnBi$_2$ has broken time-reversal symmetry\cite{Borisenko:2014}.  The simplest Hamiltonian which can describe the latter case consists of augmenting $\hat{H}_0$ to include a momentum shift $\bm{Q}$.  Similarly, the former system is modelled by adding an energy shift $Q_0$ such that the energy of the Weyl nodes is different but their position in momentum space is unchanged.  Generally, the Hamiltonian becomes\cite{Chang:2015} $\hat{H}=\hbar v_F\bm{\sigma}\cdot(\chi\bm{k}-\bm{Q})+\hat{\mathcal{I}}\chi Q_0$, where $\chi=\pm$ is the chirality of the Weyl node and $\hat{\mathcal{I}}$ is the unit matrix.  In this paper, we limit our discussion to the time-reversal symmetry breaking system but employ a more realistic Hamiltonian.  Here, the degeneracy of the Dirac cones is lifted and the two Weyl nodes sit at different locations in momentum space (although remain degenerate in energy).  

In a recent preprint, Koshino \emph{et al.}\cite{Koshino:2015} have studied the magnetic properties of 3D nodal semimetals within a three parameter model which involves the Fermi velocity, a Dirac mass parameter $m$, and an intrinsic Zeeman-field-like parameter $b$ which can arise in magnetic systems.  This model has a rich phase diagram.  A WSM exists when $b>m$ while a gapped semimetal (GSM) is present for $b<m$.  For $b=m=0$, the ordinary Dirac semimetal (DSM) is obtained; along the $m$ axis ($b=0$), the energy spectrum is gapped but the degeneracy of the Dirac cones in not lifted.  A variation of this low-energy Hamiltonian has very recently been employed by Chen \emph{et al.}\cite{Chen:2015a} in the interpretation of their magneto-infrared data in ZrTe$_5$.   Here, we will use this Hamiltonian to study the ac optical response with particular emphasis on the WSM region of the phase diagram as well as the phase boundary between the WSM and GSM. 

Our paper is organized as follows.  In Sec.~II, we introduce our low-energy Hamiltonian and discuss its phase diagram and underlying electronic dispersion.  In Sec.~III, we derive analytic equations for the ac interband background and describe the modifications from linearity brought about through the introduction of $m$ and $b$.  The details needed to compute the longitudinal conductivity are given in App.~\ref{app:A}.  Section~IV deals more specifically with the optical response at the phase boundary $m=b$ and includes a discussion of the Drude optical spectral weight.  The electronic density of states and specific heat is examined in Sec.~V.  Conclusions follow in Sec.~VI.

\section{Theoretical Formalism}

\begin{figure}[h!]
\begin{center}
\includegraphics[width=1.0\linewidth]{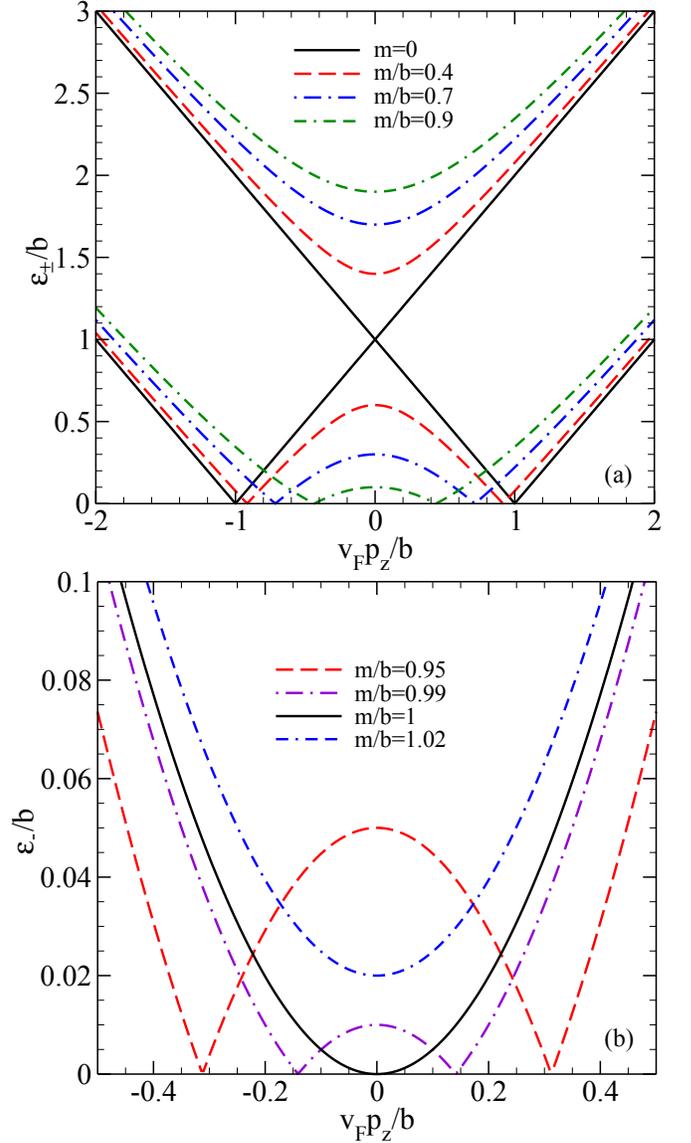}
\end{center}
\caption{\label{fig:energy}(Color online) Low-energy band structure for various values of $m/b$. (a) $s=+$ and $s^\prime=\pm$ bands for increasing $m/b$ in the WSM phase.  For finite $m$, the $s^\prime=+$ band is gapped, while $s^\prime=-$ contains two Weyl nodes. (b) $s=+$ and $s^\prime=-$ band near the phase transition ($m/b=1$). For $m/b>1$, the Weyl nodes converge and a gap emerges.
}
\end{figure}

We begin with the low-energy Hamiltonian\cite{Koshino:2015,Chen:2015a,Burkov:2011}
\begin{align}\label{Ham}
\hat{H}=\hbar v_F\hat{\tau}_x\bm{\sigma}\cdot\bm{k}+m\hat{\tau}_z+b\hat{\sigma}_z
\end{align}
where $v_F$, $m$ and $b$ are material dependent parameters.  They are the Fermi velocity, a mass parameter\cite{Chen:2015a}, and a Zeeman-like term, respectively.  Such a Hamiltonian is derived in the supplemental information of Ref.~\cite{Chen:2015a}.  Here, $\bm{\sigma}$ and $\bm{\tau}$ are two 3D vectors of Pauli matrices.  $\hat{H}$ can be cast as a 4x4 matrix, giving four energy bands (particle-hole symmetric conduction and valence bands).  The energy dispersion is
\begin{align}
\varepsilon_{ss^\prime}(\bm{p})=s\sqrt{m^2+b^2+(v_Fp)^2+2bs^\prime\sqrt{(v_Fp_z)^2+m^2}},
\end{align}
where $s=\pm$ for the conduction and valence band, respectively, $s^\prime=\pm$ and $\bm{p}=\hbar\bm{k}$ is the momentum.  For $b=0$, $\varepsilon_{ss^\prime}=s\sqrt{m^2+(v_Fp)^2}$ which is doubly-degenerate in $s^\prime$ and gives the dispersion of massive Dirac fermions. Defining $v_F\bm{p}/b\equiv\bar{\bm{p}}$ and $m/b\equiv\bar{m}$,
\begin{align}\label{energy}
\frac{\varepsilon_{ss^\prime}}{b}=s\sqrt{\bar{m}^2+1+\bar{\bm{p}}^2+2s^\prime\sqrt{\bar{p}_z^2+\bar{m}^2}}.
\end{align}
In Fig.~\ref{fig:energy}, we plot $\varepsilon_{ss^\prime}/b$ vs. $v_Fp_z/b$ at $p_x=p_y=0$.  For convenience, $s=+$ as the valence band is mirror-symmetric.
In frame (a), we consider the WSM phase with $b>m$ and four values of $\bar{m}$.  Namely: $\bar{m}=0$ (solid black), $\bar{m}=0.4$ (dashed red), $\bar{m}=0.7$ (dash-dotted blue), and $\bar{m}=0.9$ (double-dash-dotted green). For $\bar{m}=0$ (solid black), Eqn.~\eqref{energy} reduces to $\varepsilon_{ss^\prime}/b=s|1+s^\prime|\bar{p}_z||$ for which the $s^\prime=-$ branch has zeros at $\bar{p}_z=\pm 1$ and is equal to 1 at $\bar{p}_z=0$. This branch contains two Weyl nodes and an inverted cone in between. The $s^\prime=+$ has a single Dirac node but it vertically displaced in energy by one unit.  Together, these two branches can be relabled and described as two identical cones, symmetrically displaced to the right and left along the $\bar{p}_z$ axis.  For $m\neq 0$, we get $\varepsilon_{ss^\prime}/b=s|1+s^\prime\sqrt{\bar{m}^2+\bar{p}_z^2}|$ where again $p_x=p_y=0$.  This has a value of $s|1+s^\prime\bar{m}|$ at $\bar{p}_z=0$.  That is, $s(1-m/b)$ for $s^\prime=-$ and $s(1+m/b)$ for $s^\prime=+$.  The degeneracy in $s^\prime$ is lifted and two bands of the same line-type and colour are observed.  One is gapped at $\bar{p}_z=0$ ($s^\prime=+$) while the other ($s^\prime=-$) displays two nodes as a function of momentum.  Note that the separation of the nodes in $\bar{p}_z$ decreases with increasing $m$, as does the dome height at $\bar{p}_z=0$.  Specifically, the nodes are located at $\bar{p}_z=\pm\sqrt{1-m^2/b^2}$ while the height of the dome (measured from $\varepsilon/b=0$) is $1-m/b$.  Figure~\ref{fig:energy}(b) emphasizes the phase boundary at $m=b$.  Here, only $s^\prime=-$ is plotted.  For $\bar{m}<1$ the system is in the WSM phase and we see the two node structures which characterize the WSM [$\bar{m}=0.95$ (dashed red) and $\bar{m}=0.99$ (dash-dotted purple)]. For $\bar{m}=1.02$ (double-dash-dotted blue), the system is a GSM and $\varepsilon_-/b$ has a gap which tends towards zero as $\bar{m}$ tends to unity (the phase boundary).

\begin{figure}[h!]
\begin{center}
\includegraphics[width=1.0\linewidth]{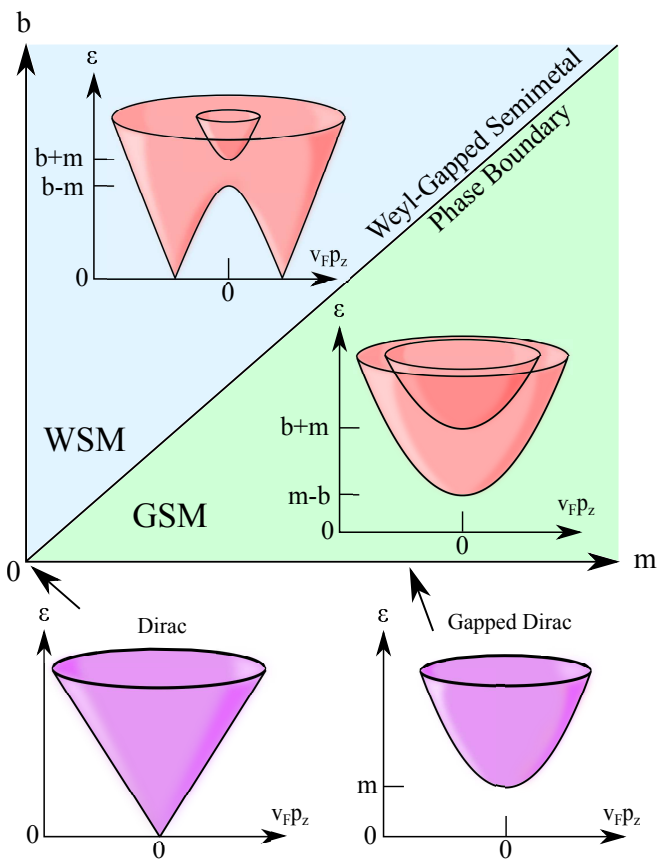}
\end{center}
\caption{\label{fig:phase}(Color online) Phase diagram of Eqn.~\eqref{Ham}.  Above $b=m$, the system is a WSM (shaded blue region).  For $b<m$ (shaded green), a GSM is present.  Along $b=0$, a degenerate gapped DSM is observed.  At $m=b=0$, degenerate massless Dirac fermions exist.  Illustrative plots of the various band structures are inset.
}
\end{figure}

The appropriate phase diagram is given in Fig.~\ref{fig:phase}.  
The phase boundary between the WSM and GSM is the line $b=m$.  Above this is the WSM phase (shaded blue).  The schematic band structure for $\varepsilon>0$ is inset.  The two nodes arise from $s^\prime=-$ while the gapped band is for $s^\prime=+$.  Below $m=b$ (shaded green) the system is a GSM and both values of $s^\prime$ display a gapped structure.  For both phases, the bands are nondegenerate.  Along $b=0$, both $s^\prime=\pm$ bands are degenerate (coloured purple) but are gapped; however at the origin ($m=b=0$), a degenerate Dirac cone exists. In all cases, $\varepsilon<0$ is mirror symmetric.

\section{Interband Conductivity}

In App.~\ref{app:A}, we derive the longitudinal optical conductivity $\sigma_{xx}(\Omega)$ via the Kubo formula in the clean limit.  Both intraband and interband transitions are considered.  The intraband component gives rise to the Drude contribution which is proportional to a Dirac $\delta$-function in the absence of impurity scattering.  The interband transitions yield a linear background as previously mentioned.  For $m=b=0$, at charge neutrality and zero temperature, it is\cite{Ashby:2014}
\begin{align}\label{IB-pure-1}
\sigma^{\rm IB}_{xx}(\Omega)=\frac{e^2}{24\pi\hbar^3v_F}\Omega,
\end{align}
for a single Dirac cone, where $e$ is the elementary charge.  The interband background is inversely proportional to the Fermi velocity, linear in $\Omega$, and zero at $\Omega=0$.  This is to be contrasted with the 2D case of graphene for which the background is universal and equal to $(\pi/2)e^2/h$.
 
For finite $m$, $b$ and $T$ the interband conductivity is [Eqn.~\eqref{IB-app}],
\begin{align}
\sigma^{\rm IB}_{xx}(\Omega)&=\frac{e^2 v_F^2\pi}{\hbar^3}\sum_{s^\prime=\pm}\int_{-\infty}^{\infty}d\omega\frac{f(\omega)-f(\omega+\Omega)}{\Omega}\notag\\
&\times \int\frac{d^3p}{(2\pi)^3}\left[1+v_F^2\frac{p_z^2-p^2}{2\varepsilon_{+s^\prime}^2}\right]\notag\\
&\times[\delta(\omega-\varepsilon_{+s^\prime})\delta(\omega+\Omega+\varepsilon_{+s^\prime})\notag\\
&+\delta(\omega+\varepsilon_{+s^\prime})\delta(\omega+\Omega-\varepsilon_{+s^\prime})],
\end{align}
where $f(\omega)=[1+{\rm exp}(\beta[\omega-\mu])]^{-1}$ is the Fermi function with chemical potential $\mu$ and $\beta=1/T$ for $k_B=1$. This gives
\begin{align}
\sigma^{\rm IB}_{xx}(\Omega)&=\frac{e^2 v_F^2\pi}{\hbar^3}\sum_{s^\prime=\pm}\int_{-\infty}^{\infty}d\omega\frac{f(\omega)-f(\omega+\Omega)}{\Omega}\delta(2\omega+\Omega)\notag\\
&\times \int\frac{d^3p}{(2\pi)^3}\left[1+v_F^2\frac{p_z^2-p^2}{2\varepsilon_{+s^\prime}^2}\right]\notag\\
&\times[\delta(\omega-\varepsilon_{+s^\prime})+\delta(\omega+\varepsilon_{+s^\prime})],
\end{align}
yielding
\begin{align}
\sigma^{\rm IB}_{xx}(\Omega)&=\frac{e^2 v_F^2\pi}{2\hbar^3}\sum_{s^\prime=\pm}\frac{f(-\Omega/2)-f(\Omega/2)}{\Omega}\notag\\
&\times \int\frac{d^3p}{(2\pi)^3}\left[1+v_F^2\frac{p_z^2-p^2}{2\varepsilon_{+s^\prime}^2}\right]\notag\\
&\times\left[\delta\left(-\frac{\Omega}{2}-\varepsilon_{+s^\prime}\right)+\delta\left(-\frac{\Omega}{2}+\varepsilon_{+s^\prime}\right)\right].
\end{align}
The thermal factor becomes
\begin{align}\label{therm}
f(-\Omega/2)-f(\Omega/2)=\frac{{\rm sinh}(\beta\Omega/2)}{{\rm cosh}(\beta\mu)+{\rm cosh}(\beta\Omega/2)}.
\end{align}
Restricting our attention to $\Omega>0$, and defining $v_F\bm{p}=\tilde{\bm{p}}$,
\begin{align}\label{IB-int}
\sigma^{\rm IB}_{xx}(\Omega)&=\frac{e^2}{16\pi^2\hbar^3 v_F}\sum_{s^\prime=\pm}\frac{1}{\Omega}\frac{{\rm sinh}(\beta\Omega/2)}{{\rm cosh}(\beta\mu)+{\rm cosh}(\beta\Omega/2)}\notag\\
&\times \int d^3\tilde{p}\left[1+\frac{\tilde{p}_z^2-\tilde{p}^2}{2\varepsilon_{+s^\prime}^2}\right]\delta\left(-\frac{\Omega}{2}+\varepsilon_{+s^\prime}\right),
\end{align}
where
\begin{align}
\varepsilon_{+s^\prime}=\sqrt{m^2+b^2+\tilde{p}^2+2bs^\prime\sqrt{\tilde{p}_z^2+m^2}}.
\end{align}
For $m=b=0$, $\varepsilon_{+s^\prime}=|\tilde{p}|$ and for $\mu=0$ and $T=0$, we get
\begin{align}\label{IB-m0b0}
\sigma^{\rm IB}_{xx}(\Omega)&=\frac{e^2}{16\pi\hbar^3 v_F}\sum_{s^\prime=\pm}\frac{1}{\Omega}\left(\frac{\Omega}{2}\right)^2\int_{-1}^1(1+{\rm cos}^2\theta)d({\rm cos}\theta)\notag\\
&=\frac{e^2}{12\pi\hbar^3v_F}\Omega,
\end{align}
which agrees with Eqn.~\eqref{IB-pure-1} when we account for an extra degeneracy of two here since we have used a four band model as opposed to the two band model used in Eqn.~\eqref{IB-pure-1}.

For $b=0$ and finite $m$, the energy dispersion is $\varepsilon_{+s^\prime}=\sqrt{m^2+\tilde{p}^2}$ which, again, is degenerate in $s^\prime$ and represents a gapped Dirac cone which remains isotropic.  For $T=0$, Eqn.~\eqref{IB-int} may now be solved to give
\begin{align}\label{IB-mb0}
\sigma^{\rm IB}_{xx}(\Omega)&=\frac{e^2}{16\pi\hbar^3 v_F}\sum_{s^\prime=\pm}\frac{1}{\Omega}\frac{\Omega}{2}\sqrt{\left(\frac{\Omega}{2}\right)^2-m^2}\left[1+\frac{2m^2}{\Omega^2}\right]\notag\\
&\times\Theta(\Omega-2m)\int_{-1}^1(1+{\rm cos}^2\theta)d({\rm cos}\theta)\notag\\
&=\frac{e^2}{6\pi\hbar^3v_F}\sqrt{\left(\frac{\Omega}{2}\right)^2-m^2}\left[1+\frac{2m^2}{\Omega^2}\right]\Theta(\Omega-2m).
\end{align}
Finite $T$ and/or $\mu$ follows by multiplying Eqn.~\eqref{IB-mb0} by Eqn.~\eqref{therm}.

The general $m$ and $b$ case is conveniently solved by using polar coordinates for $\tilde{p}_x$ and $\tilde{p}_y$ and leaving $\tilde{p}_z$ separate.  We have
\begin{align}
\sigma^{\rm IB}_{xx}(\Omega)&=\frac{e^2}{16\pi^2\hbar^3 v_F}\sum_{s^\prime=\pm}\frac{1}{\Omega}\frac{{\rm sinh}(\beta\Omega/2)}{{\rm cosh}(\beta\mu)+{\rm cosh}(\beta\Omega/2)}I_{s^\prime},
\end{align}
with
\begin{align}\label{Iz}
I_{s^\prime}&=2\int_0^\infty d\tilde{p}_z\int_0^\infty 2\pi\rho d\rho \left[2-\frac{4\rho^2}{\Omega^2}\right]\notag\\
&\times\delta\left(\frac{\Omega}{2}-\sqrt{\rho^2+(\sqrt{m^2+\tilde{p}_z^2}+s^\prime b)^2}\right).
\end{align}
This can be solved [see App.~\ref{app:B}] by introducing
\begin{align}\label{G}
G_{\pm}(x)&=\left[\Omega+\frac{4}{\Omega}(m^2+b^2)\right]x+\frac{4}{\Omega}\frac{x^3}{3}\notag\\
&\pm\frac{4b}{\Omega}\left[x\sqrt{m^2+x^2}+m^2{\rm ln}\left|\frac{x+\sqrt{x^2+m^2}}{m}\right|\right].
\end{align}
For $m>b$, which defines the GSM, 
\begin{align}\label{IB-GSM}
\sigma^{\rm IB}_{xx}(\Omega)&=\frac{e^2}{16\pi\hbar^3 v_F}\sum_{s^\prime=\pm}\frac{1}{\Omega}\frac{{\rm sinh}(\beta\Omega/2)}{{\rm cosh}(\beta\mu)+{\rm cosh}(\beta\Omega/2)}\notag\\
&\times G_{s^\prime}\left(\sqrt{\left(\frac{\Omega}{2}-s^\prime b\right)^2-m^2}\right)\Theta\left(\frac{\Omega}{2}-m-s^\prime b\right).
\end{align}
Both $s^\prime=\pm$ energy bands are gapped, one with a height of $m+b$ ($s^\prime=+$) and the other with $m-b$ ($s^\prime=-$) which is clearly illustrated in Fig.~\ref{fig:phase}. 

For the WSM ($b>m$), the $s^\prime=+$ branch is given by the same formula as Eqn.~\eqref{IB-GSM}.  For $s^\prime=-$, the function $G_-\Theta\left(\frac{\Omega}{2}-m+b\right)$ is replaced by
\begin{align}\label{Gm-WSM}
&G_-\left(\sqrt{\left(\frac{\Omega}{2}+b\right)^2-m^2}\right)\notag\\
&-G_-\left(\sqrt{\left(\frac{\Omega}{2}-b\right)^2-m^2}\right)\Theta\left(b-m-\frac{\Omega}{2}\right).
\end{align}

If $m=0$, the functional form of Eqn.~\eqref{G} reduces to
\begin{align}
G_\pm(x)=\left(\Omega+\frac{4}{\Omega}b^2\right)x+\frac{4}{3}\frac{x^3}{\Omega}\pm\frac{4b}{\Omega}x|x|.
\end{align}
For $\Omega>2b$, the conductivity involves
\begin{align}
G_+\left(-b+\frac{\Omega}{2}\right)+G_-\left(b+\frac{\Omega}{2}\right);
\end{align}
while, for $\Omega<2b$, it is proportional to
\begin{align}
G_-\left(b+\frac{\Omega}{2}\right)-G_-\left(b-\frac{\Omega}{2}\right).
\end{align}
In both cases, we get $(4/3)\Omega^2$.  Therefore, the conductivity is the same as for $m=b=0$ [Eqn.~\eqref{IB-m0b0}] as expected since we are dealing with two identical cones which are simply displaced in momentum.  This shift in $p_z$ can be eliminated through a simple change of variables in momentum space.

\begin{figure}[h!]
\begin{center}
\includegraphics[width=1.0\linewidth]{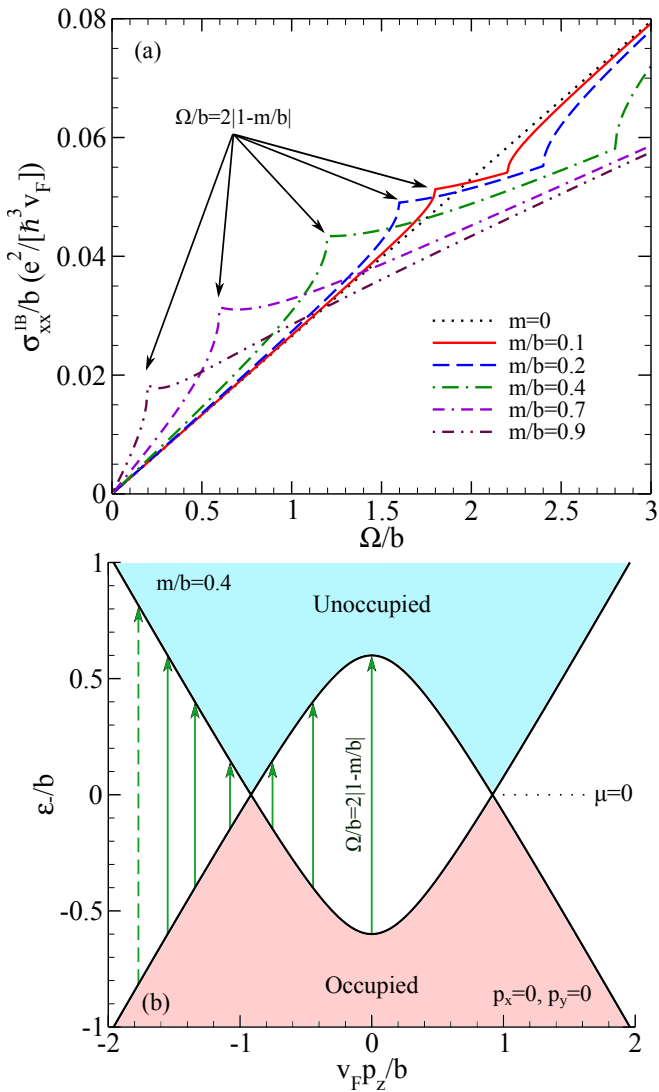}
\end{center}
\caption{\label{fig:Inter}(Color online) (a) Low-energy interband response in the WSM phase ($m/b<1$).  Two quasilinear regions are observed before the massive $s^\prime=+$ band contributes.  A van Hove singularity occurs at $\Omega/b=2(1-m/b)$. (b) $s^\prime=-$ band structure.  Characteristic optical transitions are given by the green arrows for $\mu=0$.  Above $\Omega/b=2(1-m/b)$, no transitions occur in the region between the Weyl nodes.  This yields the reduced slope seen in frame (a).
}
\end{figure}

Results for the WSM phase are shown in Fig.~\ref{fig:Inter}(a) for five values of $m/b<1$.  
The dotted black curve ($m=0$) is for comparison and is simply a straight line through the origin.  Note that all quantities have
 been scaled by $b$.  Starting with the solid red curve ($m/b=0.1$), we note a quasilinear regime extending to $\Omega/b=2(1-m/b)$ followed by a second quasilinear regime extending to $\Omega/b=2(1+m/b)$, above which a gap feature emerges and the conductivity converges with the $m=0$ curve as $\Omega\rightarrow\infty$.  While there are deviations from two perfect straight lines, they are rather minor.  This behaviour holds for the other values of $m/b$.  As this parameter is increased, the frequency regime over which the first line is seen shrinks while that over which the second line exists increases.  To understand the main features of the curves, we refer to the schematic band structure shown in frame (b).  The valence (red) and conduction (blue) bands are shown for $p_x=p_y=0$ as a function of $v_Fp_z/b$.  Arrows illustrate the various optical transitions that can occur at charge neutrality ($\mu=0$).  They range from zero to large photon energies. Note, however, that those originating from $-1<v_Fp_z/b<1$ have a maximum at $\Omega/b=2(1-m/b)$.  At this frequency, the conductivity shows a kink and rises much more slowly after this energy.  
 
It is instructive to consider how the slope of $\sigma^{\rm IB}_{xx}(\Omega)$ out of $\Omega=0$ changes with $m/b$.  We return to Eqn.~\eqref{Iz}.  Without introducing polar coordinates, we have
\begin{align}
I_{s^\prime}&=4\pi\int d\tilde{p}_xd\tilde{p}_yd\tilde{p}_z\left[2-\frac{4}{\Omega^2}(\tilde{p}_x^2+\tilde{p}_y^2)\right]\notag\\
&\times\delta\left(\frac{\Omega}{2}-\sqrt{\tilde{p}_x^2+\tilde{p}_y^2+(\sqrt{m^2+\tilde{p}_z^2}+s^\prime b)^2}\right).
\end{align}
We will restrict our attention to $s^\prime=-$ as this is the only branch which displays WSM behaviour, with two Weyl points on the $\tilde{p}_z$ axis for $\tilde{p}_x=\tilde{p}_y=0$.  As we are interested in the low-$\Omega$ dependence, only transitions in the vicinity of the Weyl nodes are present.  Therefore, we expand $\tilde{p}_z$ about the two nodes located at $\tilde{p}_{zc}=\pm\sqrt{b^2-m^2}$.  We change variables from $\tilde{p}_z$ to $\hat{p}_z\equiv\tilde{p}_z-\tilde{p}_{zc}$ and work to lowest order in $\tilde{p}_z$.  The energy of the $s^\prime=-$ branch becomes
\begin{align}
\varepsilon_{s-}\approx s\sqrt{p_x^2+p_y^2+\tilde{p}_z^2\left(1-\left(\frac{m}{b}\right)^2\right)}.
\end{align}
A change of integration variable from $\tilde{p}_z\sqrt{1-(m/b)^2}$ to $p_z$ gives an expression for the conductivity which is identical to the isotropic Dirac case with an extra factor of $1/\sqrt{1-(m/b)^2}$.  Therefore, the first quasilinear slope of the WSM interband conductivity [Fig.~\ref{fig:Inter}(a)] is altered from the dotted line ($m=0$) by a factor of $1/\sqrt{1-(m/b)^2}$.  A subtle point is that $\varepsilon_{s-}$ has two Weyl nodes and hence, a degeneracy factor of two.  The same result can be obtained from our analytic formula for the conductivity written in Eqn.~\eqref{Gm-WSM}.  This is verified in the numerical results of Fig.~\ref{fig:Inter}(a).  For $m=b\neq 0$, which defines the phase boundary between the WSM and GSM, both Eqns.~\eqref{IB-GSM} (for $m>b$) and \eqref{Gm-WSM} (for $m<b$) involve the same $G_-(\sqrt{(b+\Omega/2)^2-b^2})$. For $\Omega\rightarrow 0$, the argument of this function is $\sqrt{b\Omega}$ (to leading order) while it is $\Omega/2$ if $b=0$.  Thus, the interband conductivity at the phase boundary for small $\Omega$ follows a $\Omega^{1/2}$ law as opposed to the linear-in-$\Omega$ behaviour of $b=0$ (DSM).  We will return to this discussion in Sec.~\ref{sec:OpPhase}. 

\begin{figure}[h!]
\begin{center}
\includegraphics[width=1.0\linewidth]{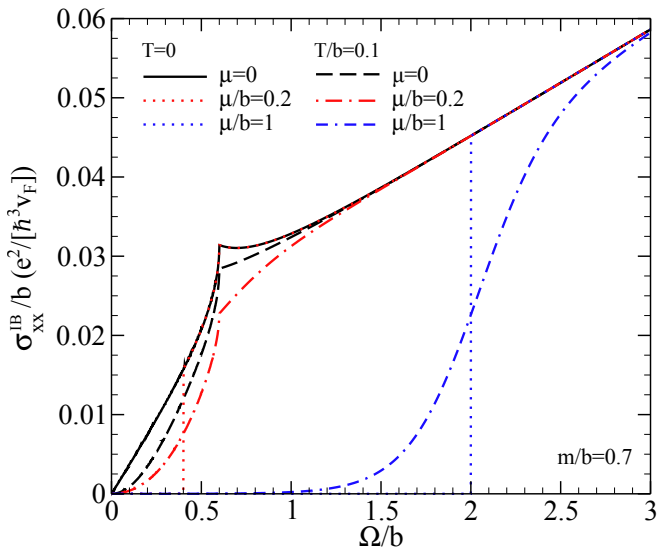}
\end{center}
\caption{\label{fig:Inter-T}(Color online) Interband optical response of the WSM for finite $T$ and $\mu$.  Finite $\mu$ causes a sharp step to occur at $\Omega=2\mu$, above which the response is identical to $\mu=0$.  Finite $T$ smears this step and a finite response is observed below $\Omega=2\mu$.
}
\end{figure}

So far, we have restricted our attention to charge neutrality and zero temperature.  In Fig.~\ref{fig:Inter-T}, we show results for the interband conductivity including finite $T$ and $\mu$ for the WSM ($m/b=0.7$).  
The solid black curve is for reference and displays the $\mu=0$, $T=0$ result.  It can be described qualitatively as two quasilinear regimes of different slope: the first (below $\Omega=2(b-m)$) going through the origin and having a slope of approximately $3.5$ times greater than the second region.  This quasilinear line extrapolates to cut the vertical axis at $\sim 0.02$ in our units. The thermal factor in Eqn.~\eqref{therm} reduces to the Heaviside step function $\Theta(\Omega-2|\mu|)$ at $T=0$.  For finite $\mu$, this cuts off the zero-$\mu$ curve at $\Omega=2|\mu|$ below which it is zero and above which it is unchanged from the charge neutral value.  This is clearly seen by the dotted red and blue curves for $\mu/b=0.2$ and 1, respectively.  The dashed black curve is for $\mu=0$ and $T/b=0.1$.  We note the depleting of the optical spectral weight at small frequencies and a decrease in the van Hove kink structure at $\Omega=2(b-m)$.  Spectral weight at finite $T$ is, of course, transferred to the low-energy Drude which is not shown in the figure.  For the dash-dotted red curve, the chemical potential is finite and cuts the low-energy region below $\Omega/b=0.4$ at $T=0$ (which falls in the first quasilinear region).  However, finite $T$ smears the step which depletes the response above $\Omega=2\mu$ and provides a finite value below.  For the dash-dotted blue curve, $\Omega/b=2\mu=2$ falls in the second quasilinear region of the optical spectrum and here the temperature smearing is more prominent.

\section{Optical Signatures of the Phase Boundary}\label{sec:OpPhase}

\begin{figure}[h!]
\begin{center}
\includegraphics[width=1.0\linewidth]{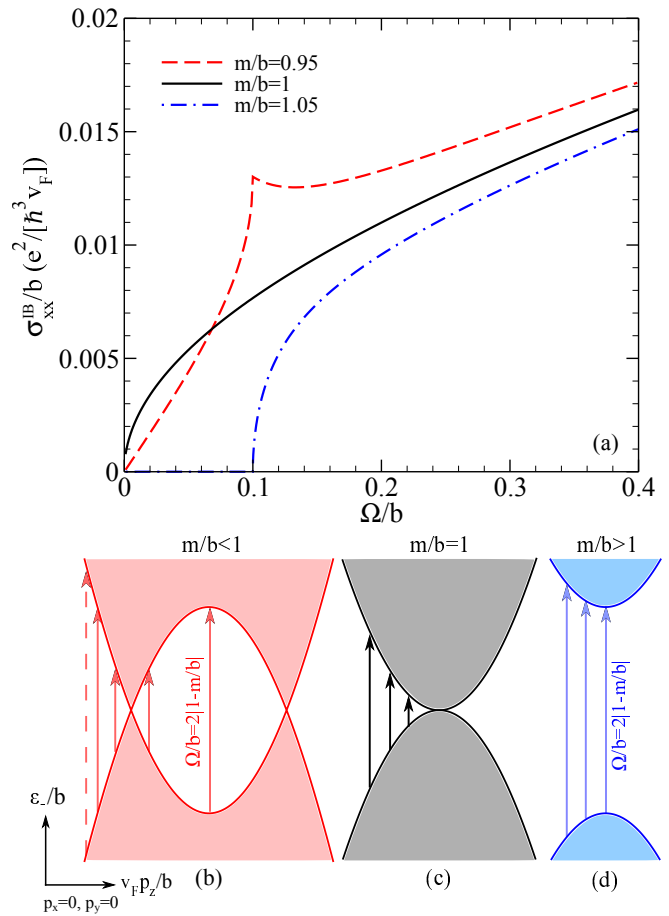}
\end{center}
\caption{\label{fig:Inter-Trans}(Color online) (a) Interband optical conductivity through the WSM to GSM phase transition.  For $m/b<1$, the system is a WSM and a quasilinear slope persists to $\Omega=0$.  At $m/b=1$, the $\Omega\rightarrow 0$ response goes like $\Omega^{1/2}$. A gap is observed for the GSM phase. Lower frame: $s^\prime=-$ band structure for the (b) WSM, (c) phase boundary, and (d) GSM.  Low energy transitions are illustrated by the arrows.
}
\end{figure}

We now turn to a closer examination of the conductivity across the WSM-GSM phase boundary ($b\approx m$). In Fig.~\ref{fig:Inter-Trans}(b)-(d), we show illustrative plots of the three important band structures for the $s^\prime=-$ energy branch as a function of $v_Fp_z/b$.  
Both valence and conduction bands are shown and key optical transitions are indicated by arrows for $\mu=0$.  Frame (b) applies to the WSM ($b>m$) and shows two Weyl nodes along the $p_z$ axis.  Frame (c) is the phase boundary ($b=m$) and the conduction and valence band touch at zero energy but are quadratic in $p_z$.  In frame (d), the GSM is shown ($b<m$) in which case the $s^\prime=-$ band is gapped.  For the WSM, optical transitions of infinitesimal energy transfer are possible around the Weyl nodes at $v_Fp_z/b=\sqrt{1-(m/b)^2}$.  The transitions about $p_z=0$ are limited to $\Omega/b=2|1-m/b|$.  For $b=m$, there is no restriction on the length of the black arrows.  For the GSM, a minimum of $\Omega/b=2|1-m/b|$ is required to excite charge carriers.  In Fig.~\ref{fig:Inter-Trans}(a), we show the interband conductivity for $m/b=0.95$ (dashed red), $m/b=1$ (solid black) and $m/b=1.05$ (dash-dotted blue). For the WSM, we see a linear low-$\Omega$ response persist to $\Omega=0$.  For the GSM, a clear gap is observed.  At high energy, the three curves converge as expected.  At the phase boundary ($b=m$), the low-$\Omega$ behaviour goes like $\sqrt{\Omega}$.  Therefore, each regime gives a very different $\Omega\rightarrow 0$ behaviour, providing a clear signature of the different phases. 

It is also useful to briefly discuss the intraband contribution (Drude) to the conductivity.  It is given by Eqn.~\eqref{Intra-app} which applies to the clean limit and can be written as
\begin{align}\label{Intra}
&\sigma^{\rm{D}}_{xx}(\Omega)=\frac{e^2}{4\pi\hbar^3 v_F}\delta(\Omega)\sum_{s,s^\prime=\pm}\int_{-\infty}^\infty d\omega\left(-\frac{\partial f(\omega)}{\partial\omega}\right)\notag\\
&\times\int\frac{d^3\tilde{p}}{2\pi}\left[\frac{\tilde{p}^2-\tilde{p}_z^2}{2\varepsilon_{ss^\prime}^2}\right]\delta(\omega-\varepsilon_{ss^\prime}).
\end{align}
The total optical spectral weight under the Drude (defined as $W_D=\int_{0^+}^\infty\sigma^{\rm{D}}_{xx}(\Omega)d\Omega$) is given by Eqn.~\eqref{Intra} with the substitution $\delta(\Omega)\rightarrow 1/2$.  For $m=b=0$, we immediately recover the known result for Dirac materials\cite{Ashby:2014},
\begin{align}
W_D=\frac{e^2}{6\pi\hbar^3v_F}\left(\mu^2+\frac{\pi^2}{3}T^2\right).
\end{align} 
For the WSM ($b>m$), we get the same answer but with a multiplicative factor of $1/\sqrt{1-(m/b)^2}$.  For $b=m\neq 0$, the $\omega^2$ factor multiplying $-\partial f/\partial\omega$ in Eqn.~\eqref{Intra} is replaced by $\sqrt{b}|\omega|^{3/2}$ (for leading order in small $\omega$) which gives
\begin{align}
W_D=\frac{e^2}{6\pi\hbar^3v_F}\sqrt{b}\int_{-\infty}^\infty d\omega\left(-\frac{\partial f(\omega)}{\partial\omega}\right)|\omega|^{3/2}.
\end{align}
This is valid for $T$ and $\mu$ small enough such that only the leading order $b|\omega|^{3/2}$ is sampled in the integral. For $T=0$ and small $\mu$,
\begin{align}
W_D=\frac{e^2}{6\pi\hbar^3v_F}\sqrt{b}|\mu|^{3/2}.
\end{align}
At charge neutrality and finite $T$, $W_D\propto T^{3/2}$.  For the GSM phase ($m>b$), the Drude is thermally activated and $W_D\propto$ exp$(-[m-b]/T)$.

\section{Density of States}

\begin{figure}[h!]
\begin{center}
\includegraphics[width=1.0\linewidth]{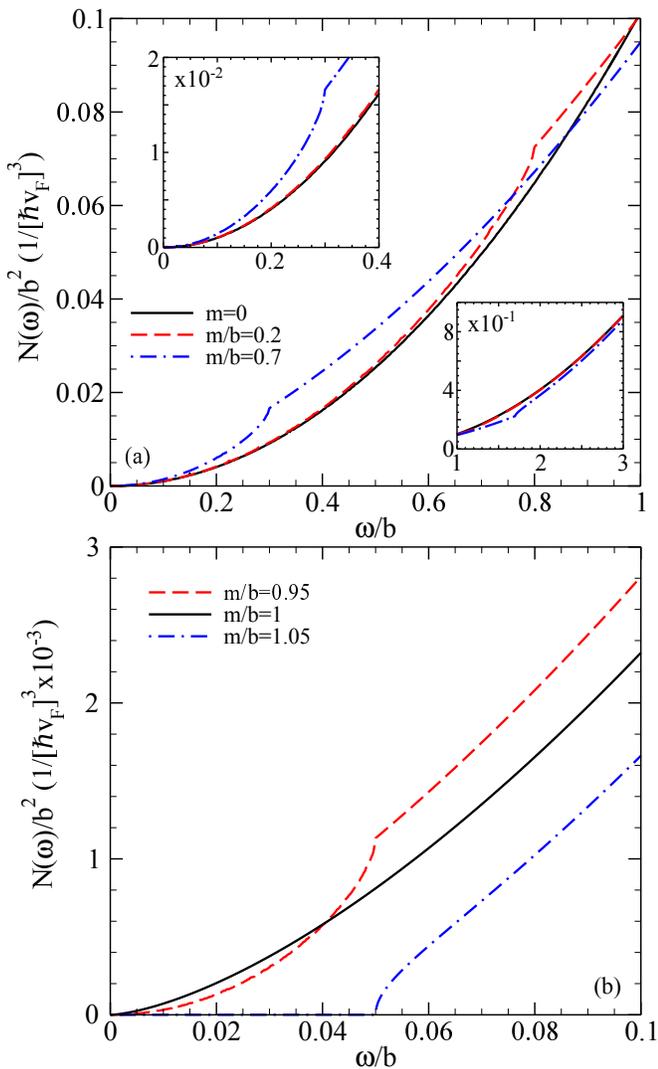}
\end{center}
\caption{\label{fig:DOS}(Color online) Low-energy DOS (a) for the WSM phase, and (b) near the phase boundary.  For the WSM, van Hove kinks are seen.  Below the first kink, the DOS grows like $\omega^2/\sqrt{1-(m/b)^2}$. For $m=b$, $N(\omega\rightarrow 0)\propto|\omega|^{3/2}$.  A low-energy gap is observed for the GSM.
}
\end{figure}

It is also instructive to look at the electronic density of states (DOS) which contains the energy conserving $\delta$-function present in the interband conductivity, but not the associated matrix element.  By definition, the DOS (in the continuum limit) is
\begin{align}
N_{s^\prime}(\omega)=\frac{1}{\hbar^3}\sum_{s}\int\frac{d^3p}{(2\pi)^3}\delta(\omega-\varepsilon_{ss^\prime}).
\end{align}
From this definition, we see that $N_{s^\prime}(\omega)=N_{s^\prime}(-\omega)$ and, hence, the DOS is particle-hole symmetric and only depends on $|\omega|$.  For $\varepsilon_{ss^\prime}=s v_F p$, we immediately get the known result\cite{Ashby:2014a}
\begin{align}
N_{s^\prime}(\omega)=\frac{\omega^2}{2\pi^2\hbar^3 v_F^3},
\end{align}
which is quadratic in $\omega$. For the GSM ($b<m$),
\begin{align}
N_\pm(\omega)=\frac{|\omega|}{2\pi^2\hbar^3v_F^3}\sqrt{\left(|\omega|\mp b\right)^2-m^2}\Theta(|\omega|\mp b-m).
\end{align}
For the WSM ($b>m$),
\begin{align}
N_+(\omega)=\frac{|\omega|}{2\pi^2\hbar^3v_F^3}\sqrt{\left(|\omega|-b\right)^2-m^2}\Theta(|\omega|- b-m),
\end{align}
while
\begin{align}\label{Nm-W}
N_-(\omega)=&\frac{|\omega|}{2\pi^2\hbar^3v_F^3}\left[\sqrt{\left(|\omega|+b\right)^2-m^2}\right.\notag\\
&\left.-\sqrt{\left(|\omega|-b\right)^2-m^2}\Theta(b-m-|\omega|)\right].
\end{align}
At the phase boundary ($b=m$), the $s^\prime=-$ branch is given by
\begin{align}
N_-(\omega)=\frac{|\omega|}{2\pi^2\hbar^3v_F^3}\sqrt{\left(|\omega|+ b\right)^2-b^2},
\end{align}
which goes as $|\omega|^{3/2}$ for $\omega\rightarrow 0$, provided $b\neq 0$. If $b$ is taken to be zero first, we find instead a $\omega^2$ power-law as expected for a DSM. The three different low-$\omega$ behaviours are shown in Fig.~\ref{fig:DOS} for several representative values of $m/b$.  
Here, $N(\omega)$ is the sum of $N_\pm(\omega)$.  Frame (a) is restricted to the WSM phase.  The solid black curve is for $m=0$ and is for comparison. In this case, $N_-(\omega)$ of Eqn.~\eqref{Nm-W} reduces to
\begin{align}
N_-(\omega)=\frac{\omega^2}{2\pi^2\hbar^3v_F^3}
\end{align}
which gives the result for the simple Dirac model.  As $m$ increases, there is little difference with the solid black curve until $\omega/b$ approaches $1-m/b$ (0.8 for $m/b=0.2$ [dashed red] and 0.3 for $m/b=0.7$ [dash-dotted blue]) where the second term in Eqn.~\eqref{Nm-W} drops out. Well below this value, the curve is quadratic, but carries an $m/b$ dependent coefficient. The $\omega\rightarrow 0$ limit is
\begin{align}
N(\omega)/b^2=\frac{(\omega/b)^2}{2\pi^2\hbar^3v_F^3}\frac{1}{\sqrt{1-(m/b)^2}}.
\end{align}
This result nicely parallels the interband conductivity.  In frame (b), we show additional results which highlight the phase transition.  Namely, $m/b=0.95$ (dashed red), $m/b=1$ (solid black) and $m/b=1.05$ (dash-dotted blue). The most significant differences between the curves occur at small $\omega/b$.  The WSM has a cusp at $\omega/b=1-m/b$ while the GSM has a gap.  The solid black curve is continuous and goes like $|\omega|^{3/2}$ at small $\omega$ while the WSM is $\propto\omega^2$ as previously noted.

The low-$\omega$ dependences of the DOS imply different temperature laws for various thermodynamic and transport properties.  Here, we illustrate this with the specific heat.  The low-$T$ electronic specific heat $\mathcal{C}(T)$ follows from the internal energy 
\begin{align}
\mathcal{U}(T)=\int_{-\infty}^\infty d\omega \omega N(\omega)f(\omega)
\end{align}
via $\mathcal{C}(T)=\partial\mathcal{U}/\partial T$ at constant volume. For the WSM\cite{Ashby:2014a},
\begin{align}
\mathcal{C}(T)\approx\frac{1}{2\pi^2\hbar^3v_F^3}\frac{b}{\sqrt{b^2-m^2}}\left(\frac{7\pi^4}{15}T^3+\pi^2\mu^2T\right),
\end{align}
for $T$ and $\mu<b-m$.  At the phase boundary for charge neutrality,
\begin{align}
\mathcal{C}(T)\approx\frac{\sqrt{b}}{2\pi^2\hbar^3v_F^3}\alpha_{7/2}T^{5/2},
\end{align}
where
\begin{align}
\alpha_{7/2}=\int_{0}^\infty\frac{x^{7/2}dx}{2{\rm cosh}^2\left(x/2\right)}.
\end{align}
For the GSM ($m>b$), the DOS is zero at small $\omega$ and thus $\mathcal{C}(T)$ becomes exponentially activated. That is, $\mathcal{C}(T)\propto$ exp$[-(m-b)/T]$, for $T<m-b$ ($m>b$).

\section{Conclusions}

We have calculated the interband optical response of relativistic Fermions within a model Hamiltonian which displays three phases. One phase contains doubly degenerate Dirac cones gapped by $m$. There is also a Weyl phase (WSM) for which the degeneracy has been lifted by a Zeeman-like magnetic parameter $b$.  Here, the two cones are centered at different points in momentum space.  Finally, there exists a semimetal phase (GSM) which is characterized by two distinct gaps.  An important aim of our work is to establish characteristics of the various phases which can be elucidated by the optical response.  For the doubly degenerate Dirac semimetal (DSM), the optical background is linear in $\Omega$ and inversely proportional to the Fermi velocity ($v_F$) of the relativistic electron dispersion.  For the WSM, the $\Omega\rightarrow 0$ behaviour remains linear in $\Omega$ but the associated $v_F$ is reduced by a factor of $\sqrt{1-(m/b)^2}$.  This region extends up to photon energies equal to $2(b-m)$ over which range it remains quasilinear.  This is followed by a second quasilinear region up to $\Omega=2(b+m)$ at which point a second gapped band begins to contribute and the resulting $\Omega$ dependence becomes more complex.  At large $\Omega$, the background converges with the linear response of the DSM.  As the boundary between the WSM and GSM is approached ($m=b$), the slope of the optical background tends toward infinity as $1/\sqrt{1-(m/b)^2}$ while, at the same time, the photon-energy range over which this law applies shrinks to zero as $2(b-m)$ ($b\geq m$).  At the phase boundary, we find a square root dependence ($\propto\Omega^{1/2}$) for the low energy response ($\Omega\rightarrow 0$).  On the other side of the phase diagram, the response is that of a gapped Dirac cone. 

For finite chemical potential ($\mu$) away from charge neutrality, the interband conductivity remains unchanged over its $\mu=0$ value except a sharp cutoff appears at $\Omega=2|\mu|$.  This holds for any value of $\mu$, independent of whether or not it falls in the first or second quasilinear region.  Finite temperature smear out this step and, consequently, introduces thermal tails below $\Omega=2|\mu|$.

We also find it instructive to calculate the density of states (DOS).  Its evolution is found to parallel the optical conductivity.  For the WSM, its $\omega\rightarrow 0$ behaviour is quadratic in $\omega$ and remains quasiquadratic until $\omega=b-m$ with its coefficient modified by the same $1/\sqrt{1-(m/b)^2}$ factor as in the optics.  At the phase boundary ($b=m$), the low-energy power-law becomes $|\omega|^{3/2}$ before a gap opens for the GSM.  This variation in the $\omega\rightarrow 0$ DOS implies changes in the low-temperature specific heat which goes from $\propto T^3$ for the WSM to $\propto T^{5/2}$ at the phase boundary, followed by an exponential activated response for the GSM phase.  Similarly, we find that the optical spectral weight of the Drude conductivity in the WSM phase is the same as the Dirac semimetal up to a factor of $1/\sqrt{1-(m/b)^2}$ for $T<b-m$.  For $b=m$, the Drude weight goes like $|\mu|^{3/2}$ at $T=0$, and $T^{3/2}$ at charge neutrality.  This is distinct from both the WSM and GSM.

\begin{acknowledgments}
This work has been supported by the Natural Sciences and Engineering Research Council (NSERC) (Canada) and, in part, by the Canadian Institute for Advanced Research (Canada).
\end{acknowledgments}

\appendix
\section{}\label{app:A}

Working in the one-loop approximation, the real part of the Kubo formula at finite frequency is
\begin{align}\label{Kubo-app}
&\sigma_{\alpha\beta}(\Omega)=\frac{e^2\pi}{\hbar^3\Omega}\int_{-\infty}^\infty d\omega[f(\omega)-f(\omega+\Omega)]\notag\\
&\times\int\frac{d^3p}{(2\pi)^3}{\rm Tr}[\hat{v}_\alpha\hat{\mathcal{A}}(\bm{p},\omega)\hat{v}_\beta\hat{\mathcal{A}}(\bm{p},\omega+\Omega)],
\end{align}
where $f(\omega)=[{\rm exp}(\beta[\omega-\mu]+1]^{-1}$ is the Fermi function, with $\beta=1/T$.  The velocity operator ($\hat{v}_\alpha$) is related to the Hamiltonian ($\hat{H}$) through $\hat{v}_\alpha=\partial\hat{H}/\partial p_\alpha$.  The spectral function $\hat{\mathcal{A}}$ is related to the Green's function by
\begin{align}
\hat{\mathcal{G}}(\bm{p},z)=\int_{-\infty}^\infty\frac{\hat{\mathcal{A}}(\bm{p},\omega)}{z-\omega}d\omega,
\end{align}
where
\begin{align}
\hat{\mathcal{G}}^{-1}(\bm{p},z)=\hat{I}z-\hat{H},
\end{align}
with $\hat{I}$ the identity matrix of dim($\hat{H}$).  Using Eqn.~\eqref{Ham} and considering the longitudinal conductivity ($\alpha=\beta=x$), the necessary spectral elements ($\mathcal{A}_{ij}$) of the spectral matrix are
\begin{align}
\mathcal{A}_{11}=\sum_{s,s^\prime=\pm}\alpha_{ss^\prime}\delta(\omega-\varepsilon_{ss^\prime}),
\end{align}
\begin{align}
\mathcal{A}_{12}=\sum_{s,s^\prime=\pm}\beta_{ss^\prime}\delta(\omega-\varepsilon_{ss^\prime}),
\end{align}
\begin{align}
\mathcal{A}_{13}=\sum_{s,s^\prime=\pm}\gamma_{ss^\prime}\delta(\omega-\varepsilon_{ss^\prime}),
\end{align}
and
\begin{align}
\mathcal{A}_{14}=\sum_{s,s^\prime=\pm}\delta_{ss^\prime}\delta(\omega-\varepsilon_{ss^\prime}),
\end{align}
where $\varepsilon_{ss^\prime}$ is given by Eqn.~\eqref{energy} and
\begin{align}
&\alpha_{ss^\prime}=\frac{s^\prime}{2\varepsilon_{ss^\prime}[\varepsilon_{++}^2-\varepsilon_{+-}^2]}\left(\varepsilon_{ss^\prime}^3+\varepsilon_{ss^\prime}^2(m+b)+2bv_F^2p_z^2\right.\notag\\
&\left.-\varepsilon_{ss^\prime}[(m-b)^2+v_F^2p^2]-(m+b)[(m-b)^2+v_F^2p^2]\right),
\end{align}
\begin{align}
\beta_{ss^\prime}=s^\prime\frac{bv_F^2p_z(p_x-ip_y)}{\varepsilon_{ss^\prime}[\varepsilon_{++}^2-\varepsilon_{+-}^2]},
\end{align}
\begin{align}
&\gamma_{ss^\prime}=\frac{s^\prime v_F p_z}{2\varepsilon_{ss^\prime}[\varepsilon_{++}^2-\varepsilon_{+-}^2]}\left(\varepsilon_{ss^\prime}^2+2b\varepsilon_{ss^\prime}+b^2-m^2-v_F^2p^2\right),
\end{align}
and
\begin{align}
&\delta_{ss^\prime}=\frac{s^\prime(p_x-ip_y)}{2\varepsilon_{ss^\prime}[\varepsilon_{++}^2-\varepsilon_{+-}^2]}\left(\varepsilon_{ss^\prime}^2-(m-b)^2-v_F^2p_z^2\right).
\end{align}
All other elements can be related to these four.  Evaluating the trace, the intraband contribution to the conductivity is
\begin{align}\label{Intra-app}
&\sigma^{\rm{D}}_{xx}(\Omega)=\frac{e^2v_F^2\pi}{\hbar^3\Omega}\sum_{s,s^\prime=\pm}\int_{-\infty}^\infty d\omega[f(\omega)-f(\omega+\Omega)]\notag\\
&\times\int\frac{d^3p}{(2\pi)^3}\left[\frac{v_F^2(p^2-p_z^2)}{2\varepsilon_{ss^\prime}^2}\right]\delta(\omega-\varepsilon_{ss^\prime})\delta(\omega+\Omega-\varepsilon_{ss^\prime}).
\end{align}
The interband component is
\begin{align}\label{IB-app}
&\sigma^{\rm{IB}}_{xx}(\Omega)=\frac{e^2v_F^2\pi}{\hbar^3\Omega}\sum_{s,s^\prime=\pm}\int_{-\infty}^\infty d\omega[f(\omega)-f(\omega+\Omega)]\notag\\
&\times\int\frac{d^3p}{(2\pi)^3}\left[1+\frac{v_F^2(p_z^2-p^2)}{2\varepsilon_{ss^\prime}^2}\right]\delta(\omega-\varepsilon_{ss^\prime})\delta(\omega+\Omega+\varepsilon_{ss^\prime}).
\end{align}
To solve this numerically, it is convenient to conduct the $d^3p$ integral in spherical coordinates and define $\epsilon=v_F p$.

\section{}\label{app:B}

To obtain an analytic expression for the interband conductivity, we need to evaluate Eqn.~\eqref{Iz}:
\begin{align}\label{Iz-app}
I_\pm&=2\int_0^\infty d\tilde{p}_z\int_0^\infty 2\pi\rho d\rho \left[2-\frac{4\rho^2}{\Omega^2}\right]\notag\\
&\times\delta\left(\frac{\Omega}{2}-\sqrt{\rho^2+(\sqrt{m^2+\tilde{p}_z^2}\pm b)^2}\right).
\end{align}
To begin, define
\begin{align}
a_\pm=\sqrt{m^2+\tilde{p}_z^2}\pm b. 
\end{align}
The Dirac $\delta$-function in Eqn.~\eqref{Iz-app} only clicks for $\Omega>2|a_\pm|$.  Defining $\epsilon=\rho^2$,
\begin{align}
I_\pm&=2\pi\int_0^\infty d\tilde{p}_z\int_0^\infty  d\epsilon \left[2-\frac{4\epsilon}{\Omega^2}\right]\delta\left(\frac{\Omega}{2}-\sqrt{\epsilon+a_\pm^2}\right),
\end{align}
which we recast as
\begin{align}
I_\pm &=2\pi\int_0^\infty d\tilde{p}_z\int_0^\infty  2EdE \left[1+\frac{4}{\Omega^2}a_\pm^2\right]\delta\left(\frac{\Omega}{2}-E\right).
\end{align}
This simplifies to
\begin{align}\label{Iz-G}
I_\pm&=2\pi\int_0^\infty d\tilde{p}_z\Omega\left[1+\frac{4}{\Omega^2}a_\pm^2\right]\notag\\
&\equiv 2\pi\int_0^\infty d\tilde{p}_z\frac{dG_\pm}{d\tilde{p}_z},
\end{align}
with the condition $|a_\pm|<\Omega/2$, where
\begin{align}\label{G-app}
G_{\pm}(\tilde{p}_z)&=\left[\Omega+\frac{4}{\Omega}(m^2+b^2)\right]\tilde{p}_z+\frac{4}{\Omega}\frac{\tilde{p}_z^3}{3}\notag\\
&\pm\frac{4b}{\Omega}\left[\tilde{p}_z\sqrt{m^2+\tilde{p}_z^2}+m^2{\rm ln}\left|\frac{\tilde{p}_z+\sqrt{\tilde{p}_z^2+m^2}}{m}\right|\right].
\end{align}

Consider first the case of $m>b$.  The condition $\Omega>2|a_\pm|$ is simply $\Omega>2(\sqrt{m^2+\tilde{p}_z^2}\pm b)$, or
\begin{align}
\tilde{p}_z<\sqrt{\left(\frac{\Omega}{2}\mp b\right)^2-m^2},
\end{align}
which puts a condition on the upper limit in the integral of Eqn.~\eqref{Iz-G} and implies $\Omega/2>m\pm b$ which is always satisfied for $m>b$.  Thus,
\begin{align}
I_\pm &=2\pi\int_0^{\sqrt{(\Omega/2-b)^2-m^2}}\frac{dG_\pm}{d\tilde{p}_z}d\tilde{p}_z\Theta\left(\frac{\Omega}{2}-m\mp b\right).
\end{align}
Finally, we arrive at
\begin{align}
I_\pm=2\pi G_\pm\left(\sqrt{\left(\frac{\Omega}{2}\mp b\right)^2-m^2}\right)\Theta\left(\frac{\Omega}{2}-m\mp b\right).
\end{align}

The WSM case ($b>m$) requires more care since the negative branch $a_-=\sqrt{\tilde{p}_z^2+m^2}-b$ is not always positive.  It, in fact, changes sign as a function of $\tilde{p}_z$.  For $\sqrt{\tilde{p}_z^2+m^2}>b$, $a_->0$; however, $a_-<0$ in the opposite case.  We also need to apply the condition $\Omega/2>|a_\pm|$ which gives $\Omega/2>\sqrt{\tilde{p}_z^2+m^2}-b$ in the first instance and $\Omega/2>b-\sqrt{\tilde{p}_z^2+m^2}$ in the latter.  Thus for $a_->0$, we have two conditions on the $\tilde{p}_z$ integration.  Namely, $\sqrt{b^2-m^2}<\tilde{p}_z<\sqrt{(\Omega/2-b)^2-m^2}$.  For $a_-<0$, the conditions are $\sqrt{(\Omega/2-b)^2-m^2}<\tilde{p}_z<\sqrt{b^2-m^2}$ and, of course, $-\Omega/2+b>m$.  We get
\begin{align}
I_-=2\pi &\left[G_-\left(\sqrt{\left(\frac{\Omega}{2}+b\right)^2-m^2}\right)\right.\notag\\
&\left.-G_-\left(\sqrt{\left(\frac{\Omega}{2}-b\right)^2-m^2}\right)\Theta\left(b-m-\frac{\Omega}{2}\right)\right].
\end{align}

We have verified the approach to $b=m$ from the WSM and GSM phases agree.  In that case, the $\Omega\rightarrow 0$ limit of $G_-$ is particularly simple and reduces to $\sqrt{b}\Omega^{3/2}$.

\bibliographystyle{apsrev4-1}
\bibliography{WD}

\end{document}